\documentstyle[pic]{article}

\let\citep\cite

\begin{document}

\title{Why Physics Needs Quantum Foundations}
\author{Lucien Hardy and Robert Spekkens}
\address{Perimeter Institute for Theoretical Physics, 31 Caroline St. N, Waterloo,
Ontario, Canada N2L 2Y5}
\maketitle

Quantum theory is a peculiar creature. It was born as a theory of atomic
physics early in the twentieth century, but over time its scope has
broadened, to the point where it now underpins all of modern physics with the
exception of gravity. It has been verified to extremely
high accuracy and has never been contradicted experimentally. Yet despite
its enormous success, there is still no consensus among physicists about
what this theory is saying about the nature of reality. There is no question that quantum theory
works well as a tool for predicting what will occur in experiments. But
just as understanding how to drive an automobile is different from understanding how it works or how to fix it should it break down, so too is there a difference between understanding how
to use quantum theory and understanding what it means. The field of
quantum foundations seeks to achieve such an understanding. In particular,
it seeks to determine the correct interpretation of the formalism. It also
seeks to determine the principles that underlie quantum theory. Why do we
have a quantum world as opposed to a classical world or some other kind of
world entirely?

There are many motivations for pursuing foundational research. One is the
development of quantum technologies, such as quantum computation and quantum
cryptography. A better understanding of the
theory facilitates the identification and development of these new technologies,
the harnessing of the power of nonclassicality.   Another motivation is that quantum theory is likely not the end of the road.  If we are to move beyond it, then it
is important to know which parts can be changed or generalized or abandoned.  Finally, there are the personal motivations of individual researchers: because
quantum theory is very mysterious and counterintuitive and surprising and it
seems to defy us to understand it. And so we take up the challenge.

\subsection{Operationalism and realism}

Broadly speaking, researchers in quantum foundations can be divided into
two camps. There are the operationalists and there are the realists. For the
operationalist, operators in Hilbert space represent preparation and
measurement procedures, specified as lists of
instructions of what to do in the lab.  They are recipes with macroscopic activities
as ingredients. The theory merely specifies what probabilities of
outcomes will be observed when a given measurement follows a given
preparation. For the realist, there is some deeper reality underlying the
equations of quantum theory that ultimately accounts for why we see the
relative frequencies we do. For the realist, quantum theory needs an
interpretation. Does the wave function describe a real entity? Are there
extra hidden variables in addition to the wave function needed to fully
describe a quantum system?

A classic example of the power of applying operational thinking is Einstein's approach to special relativity.  By carefully considering how to synchronise distant clocks, he was led to abandon the hitherto cherished notion of absolute simultaneity.  A good example of the successful application of realism is the atomic hypothesis.  In this case, John Dalton and others were right to insist on the reality of atoms (in opposition to operationalists such as Ernst Mach).  It led to a theory for Brownian motion(Einstein again), the theory of statistical mechanics, and ultimately much of modern physics.

Historically, both approaches were important in the development of quantum theory. Heisenberg's 1925 paper on matrix mechanics, which ushered in the modern age of quantum theory, began with the sentence ``The present paper seeks to establish a basis for theoretical
quantum mechanics founded exclusively upon relationships between quantities
which in principle are observable." This was operational thinking. In
parallel to this, de Broglie posited the existence of waves to describe
quantum phenomena and Schr\"{o}dinger found an equation for their motion. This
was realist thinking.

In modern research into the foundations of quantum theory, both operationalism and realism are alive and well.  By thinking operationally, a general mathematical framework has been developed which can accommodate a wide variety of probabilistic theories. Quantum theory fits very comfortably into this framework as a special case and so can be easily understood in operational
terms. Much progress has been made recently in understanding the deeper
mathematical structure of quantum theory in the context of this mathematics
of operationalism. For example, many features of
quantum theory (such as the impossibility of building a machine that can clone quantum states) turn out to be features of any non-classical probabilistic theory.
These tools also contribute to the program of reconstructing quantum theory deriving its abstract mathematical structure from natural postulates, just as the Lorentz transformations
are derived from Einstein's postulates for special relativity.

But operationalism is not enough. Explanations do not bottom out with
detectors going `click'. Rather, the existence of detectors that click is
the sort of thing that we can and should look to science to explain. Indeed,
science seeks to explain far more than this, such as the existence of human
agents to build these detectors, the existence of an earth and a sun to
support these agents, and so on to the existence of the universe itself.
The only way to meet these challenges
is if explanations do not bottom out with complex entities and everyday
concepts, but rather with simple entities and abstract concepts. This is the
view of the realist. Without adopting some form of realism, it is unclear
how one can seek a complete scientific world-view, incorporating not just
laboratory physics, but all scientific disciplines, from evolutionary
biology to cosmology. It is true of course that all of our evidence will
come to us in the form of macroscopically observable phenomena, but we need
not and should not restrict ourselves to these concepts when constructing
scientific theories. For the realist, then, we need an interpretation of
quantum theory.

There are already plenty of candidates to choose from. There is the pilot
wave model of Louis de Broglie and David Bohm in which the wavefunction
guides the motion of actual particles according to a well defined equation.
There is the many worlds interpretation of Hugh Everett III in which the
universe is regarded as splitting into many copies every time the
wavefunction evolves into a superposition of distinct situations. There are
also collapse models in which extra terms are added to the Schr\"{o}dinger
equation to cause a collapse of the wavefunction when sufficiently macroscopic possibilities become superposed. Many more ideas for interpretations are in the making today. Cases have been made for each by their respective proponents, but none has yet proven sufficiently compelling to achieve a scientific consensus.  So research on these issues continues.

Ultimately, we expect that both operationalism and realism will
play an important methodological role in future research. Operationalism is,
at least, a useful exercise for freeing the mind from the baggage of
preconceptions about the world, as Einstein did when he showed that
the notion of absolute simultaneity was unfounded. As such it can provide a
minimal interpretation, some conceptual and mathematical scaffolding on
which to build.  On the other hand, the extra commitments, constraints and details of a realist model can also be a virtue. Realist models are more falsifiable, they typically suggest new and interesting questions (questions that may uncover novel consequences of a theory), and they often suggest avenues for modifying and generalizing the theory.

\subsection{The foundational roots of quantum information theory}

The field of quantum foundations provides many examples of how basic
research guided by a desire for deeper understanding can lead to discoveries
of great practical interest. Quantum information science serves as the
best example.  To first approximation, it was born of two communities: on the one hand, computer scientists and information theorists, and on the other, physicists
thinking about the foundations of quantum theory. If
the name of a field indicated its parentage, then the ``Quantum'' in
``Quantum Information'' would refer to Quantum Foundations.

Since those early days, there has been a slow but steady march towards quantum technologies becoming practical. Quantum cryptographic systems, for instance, are now available commercially.  Meanwhile, progress on the theoretical side has shown how one can achieve stronger forms of
security than previously conceived.  One of the most
celebrated cryptographic applications of quantum theory is key distribution:
the ability to establish a shared secret key among distant parties over a
public channel in such a way that one can reliably detect the presence of an
eavesdropper.  Recent work
has shown that under the very conservative assumption that superluminal signalling is
impossible, one can achieve key distribution even if the would-be
eavesdropper has the advantage of providing the very devices that are used
by the communicating parties.\cite{BHK,AGM}

This is practical stuff, but the path that led to such results starts with
foundational research.  In 1964, John Bell was considering the
question of whether there is an interpretation of quantum theory in terms of
hidden variables. He had been pondering the argument by Einstein, Podolsky
and Rosen in favor of the incompleteness of the quantum description and
thinking about various theorems that purported to show the impossibility of
such completions. He was also studying the pilot wave model of deBroglie
and Bohm. He noted that this theory postulated superluminal causal
influences and wondered whether this might be true of all realist models of
quantum theory. Once the question was asked, it was not long before he was
able to prove that this is indeed the case -- a theorem that now bears his
name.\cite{Bell}

Bell's theorem is a profound result because it demonstrates a tension
between the two pillars upon which modern physics is built -- quantum
physics and relativity theory. Since its discovery, physicists have been
puzzling over it. One such person was Artur Ekert. In
1991, he realized that the statistical correlations central to Bell's theorem
could be used to achieve secure key distribution.\cite{Ekert} Although a different
quantum protocol for key distribution had been developed seven years earlier
by Charlie Bennett and Gilles Brassard\cite{BB84}, it was Ekert's protocol that
ultimately led to the results mentioned above -- the possibility of
achieving security regardless of the provenance of the devices.

The theory of entanglement -- the property of quantum states that is critical to the Einstein-Podolsky-Rosen argument and Bell's theorem -- is another example of the practical payoff of foundational thinking. In 1980, William Wootters had just completed a Ph.D. thesis on a foundational question: from what principles can the Born rule of quantum theory be derived? Important to his
considerations was a task known as quantum state tomography. This is an attempt to infer the identity of a quantum state by implementing many different measurements on a large number of samples of it.
In the fall of 1989, Asher Peres, another foundational researcher, asked whether
\emph{joint} measurements on a pair of systems might yield better tomography than separate measurements.  They were able to find strong numerical evidence that
this was indeed the case.\cite{PeresWootters} It seemed, therefore, that if a pair of similarly prepared
particles was separated in space, an experimenter would be less able to identify their state
than if they were together. In other words, there is a limit to how much
information about the state can be accessed by local means -- a kind of
nonlocality. In 1992, Charlie Bennett heard a talk by Wootters on the
subject and asked whether the nonlocality that seemed to be
inherent in entangled states might provide a way of achieving state tomography on separated systems with the same success that could be achieved if they were proximate.

Again, once the question was asked, it took only a few days for Wootters,
Bennett, Peres and their co-workers (Gilles Brassard, Claude Cr\'{e}peau and Richard Jozsa) to answer it. Yes, it could be done.\cite{teleportation} The key insight was that by consuming a maximally entangled state (i.e. using it in a manner that ultimately destroys it), the quantum state of a system could be transferred from one party to another distant party using only local operations and classical communication.  The trick was dubbed ``quantum teleportation'' by its authors.  Several discoveries in quantum information theory (including Ekert's key distribution protocol) had shown that entanglement was useful, and with the discovery of teleportation, it became especially obvious: entanglement was a resource.  This change in perspective
prompted researchers to ask many new and interesting questions about
entanglement.  The result has been a dramatic increase in our understanding of
the phenomena, leading to applications across all subdisciplines of quantum
information science (cryptography, communication and computation) and
further afield (for instance, in new density matrix renormalization group methods for
simulating quantum many-body systems).

One final story. Early in the history of quantum information theory, when
most researchers were still thinking about quantum theory as imposing upon
us additional limitations relative to what we would face in a world that was
governed by a classical theory, David Deutsch was thinking differently.  He
was looking to identify tasks for which quantum theory provided an
advantage.  In the mid-eighties, his unique perspective led him to write
one of the very first articles on quantum computation, an article that
prepared the ground for important subsequent discoveries.\cite{Deutsch}  What led Deutsch
to perform this seminal work?  He was thinking about the
information-processing consequences of Everett's many worlds interpretation
of quantum theory.

\subsection{Quantum foundations meets quantum gravity}

Perhaps the holy grail of modern physics is a theory of quantum gravity. We need to find a theory that reduces to quantum theory in one limit
and to general relativity in another, and that makes new
predictions which are subsequently verified in experiments. This
has been an open problem since the birth of quantum theory, yet we still do not
have a theory of quantum gravity. The problem is difficult because
there are deep conceptual differences between general relativity and quantum
theory. Consequently, the two theories have very different mathematical
structures.

In the past, when two less fundamental theories have been unified into a deeper, more fundamental theory, the unification has typically required an entirely new mathematical framework, motivated by conceptual insights from the two component theories.
If this is the case for quantum gravity, then foundational thinking is likely to be useful. Does
quantum gravity call for a new type of probabilistic theory? Which of the
postulates of quantum theory (in whatever formulation) will have to be
modified or abandoned, if any? A similar type of conceptual thinking about
the foundations of general relativity is also likely to be significant. If
we have a mathematical framework that is rich enough to contain a
theory of quantum gravity (in much the same way that the mathematics of
Hilbert space is sufficient for quantum theory and the mathematics of tensor
calculus is sufficient for general relativity) then we could expect that a
few suitably chosen postulates would narrow us down to the right theory. It is in the construction of this framework and the selection of these postulates that the conceptual and mathematical tools of quantum
foundations are likely to be useful.

\subsection{Send off}

The field of quantum foundations does not merely exist to tidy up the mess
left behind after the physics has been done. Rather it should be regarded as
part and parcel of the great project of theoretical physics - to gain an
ever better understanding of the world around us.

In particular, researchers in the field are striving to achieve a deeper understanding of the conceptual and mathematical structure of quantum theory.
It is a testament to the importance of this sort of pure enquiry that the ideas of quantum foundations have found such a compelling application in the field of quantum information science.   It was John Bell thinking about hidden variables that ultimately led to many practical results in quantum cryptography; it was William Wootters asking ``Why the Born rule?'' that guided us down the last stretch of the path that culminated in understanding entanglement as a resource, and it was David Deutsch thinking about the many worlds interpretation of quantum theory that laid the foundations of quantum computing.

We should not expect that quantum information theory will be the only substantial application of ideas from quantum foundations. They may well play a significant role in the construction of a theory of quantum gravity.  They may even spawn entirely new fields of research that we cannot currently predict. Thinking about foundations pays off in the long run.  David Mermin once summarized a popular attitude towards quantum theory as ``Shut up and calculate!''. We suggest an alternative slogan: ``Shut up and contemplate!''

\section*{Acknowledgements}
Research at Perimeter Institute is supported by the Government of Canada through Industry Canada and by the Province of Ontario through the Ministry of Research and Innovation.


\begin{thebibliography}{9}

\bibitem{BHK} J. Barrett, L. Hardy, and A. Kent, \emph{No Signaling and Quantum Key Distribution}, Phys. Rev. Lett. \textbf{95}, 010503 (2005).
\bibitem{AGM} A. Ac\'{\i}n, N. Gisin, Ll. Masanes, \emph{From Bell's Theorem to Secure Quantum Key Distribution}, Phys. Rev. Lett. \textbf{97}, 120405 (2006).
\bibitem{Bell} J. S. Bell, \emph{On the Einstein-Podolsky-Rosen paradox}, Physics (Long Island City, N.Y.) \textbf{1}, 195 (1964).
\bibitem{Ekert} A. K. Ekert, \emph{Quantum cryptography based on Bell's theorem}, Phys. Rev. Lett. \textbf{67}, 661 (1991).
\bibitem{BB84} C. H. Bennett and G. Brassard, \emph{Quantum Cryptography: Public key distribution and coin tossing}, in Proceedings of the IEEE International Conference on Computers, Systems, and Signal Processing, Bangalore, p. 175 (1984).
\bibitem{PeresWootters} A. Peres and W. K. Wootters, \emph{Optimal detection of quantum information}, Phys. Rev. Lett. \textbf{66}, 1119 (1990).
\bibitem{teleportation} C. H. Bennett, G. Brassard, C. Crépeau, R. Jozsa, A. Peres, W. K. Wootters, \emph{Teleporting an Unknown Quantum State via Dual Classical and Einstein-Podolsky-Rosen Channels}, Phys. Rev. Lett. \textbf{70}, 1895 (1993).
\bibitem{Deutsch} D. Deutsch, \emph{Quantum theory, the Church-Turing principle and the universal quantum computer}, Proc. R. Soc. Lond. A \textbf{400} 97 (1985).
\end{thebibliography}
\end{document}